\documentclass[conference]{IEEEtran}
\IEEEoverridecommandlockouts
\usepackage{cite}
\usepackage{amsmath,amssymb,amsfonts}
\usepackage{algorithmic}
\usepackage{subcaption}

\usepackage{graphicx}
\usepackage{textcomp}
\usepackage{xcolor}
\usepackage{ulem}

\usepackage[para]{threeparttable}

\def\BibTeX{{\rm B\kern-.05em{\sc i\kern-.025em b}\kern-.08em
    T\kern-.1667em\lower.7ex\hbox{E}\kern-.125emX}}
\begin{document}

\title{LUCAS: A Low-Power Ultra-Low Jitter Compact ASIC for SiPM Targetting ToF-CT}

\author{
\IEEEauthorblockN{Seyed Arash Katourani\IEEEauthorrefmark{1}\IEEEauthorrefmark{2}, Marwan Besrour\IEEEauthorrefmark{1}\IEEEauthorrefmark{2}, Takwa Omrani\IEEEauthorrefmark{1}\IEEEauthorrefmark{2}, Konin Koua\IEEEauthorrefmark{1}\IEEEauthorrefmark{2}, Maher Benhouria\IEEEauthorrefmark{1}\IEEEauthorrefmark{2},\\ Gianluca Giustolisi\IEEEauthorrefmark{3}, Réjean Fontaine%
\IEEEauthorrefmark{1}\IEEEauthorrefmark{2}, and Marc-André Tétrault%
\IEEEauthorrefmark{1}\IEEEauthorrefmark{2}}
\IEEEauthorblockA{\IEEEauthorrefmark{1}\textit{Université de Sherbrooke}}
\IEEEauthorblockA{\IEEEauthorrefmark{2}\textit{Interdisciplinary Institute for Technological Innovation (3IT)}}
\IEEEauthorblockA{\IEEEauthorrefmark{3}\textit{Universit\`a degli Studi di Catania, Catania, Italy}}
}
\maketitle

\begin{abstract}
We present LUCAS (\underline{L}ow power \underline{U}ltra-low jitter \underline{C}ompact \underline{A}SIC for \underline{S}iPM), an analog front-end for Silicon Photomultipliers (SiPM) targeting fast timing detectors in Time-of-Flight Computed Tomography (ToF-CT). LUCAS features a very low input impedance preamplifier followed by a voltage comparator. It is designed in TSMC 65\,nm low-power CMOS technology with a power supply of 1.2~V. Our first 8-channel prototype has been sent to fabrication and will be received in August 2023.

Post-layout simulations 
predict less than 40\,ps FWHM SPTR jitter and an approximate power consumption of 3.2\,mW per channel. 
The front end is suitable for applications with rigorous jitter requirements and high event rates, thanks to its 3.9\,GHz unity-gain bandwidth. The front-end compact form factor will facilitate its incorporation into systems demanding high channel densities.


\end{abstract}

\section{Introduction}
SiPMs have become a key technology for compact, pixelated, fast scintillating detectors dedicated to time-of-flight (ToF) measurement applications, particularly in medical imaging. SiPM consists of an array of single-photon avalanche diodes (SPADs) which operates above the breakdown voltage, providing exceptional photon sensitivity and current amplification through an avalanche process. To effectively utilize this phenomenon, a low input impedance amplifier is crucial for the SiPM front end readout. Lower input impedance allows better avalanche current management and addresses the SiPM inherent parasitic capacitance, preventing a low-frequency node at the input stage. Low input impedance not only improves the received current amplitude but also enables faster detection, enhancing the overall performance of the target applications.

Reducing the input impedance has been extensively studied with different strategies like cross-coupled current conveyors~\cite{b5} (with instability risks), regulated cascade and current reuse amplifiers~\cite{b9} (high power consumption), and transimpedance amplifiers~\cite{TIA} (low DC gain). Moreover, recent efforts to reduce SiPM parasitic capacitance using balun transformers have been effective~\cite{BALUN}; however, their implementation requires additional space, limiting channel density increase. 

To address the specified issue, we introduce a sophisticated front-end system that concurrently targets outstanding timing performance, diminished power consumption, and a compact design, accomplished through the application of advanced analog circuit methodologies and the utilization of lower node technology. The adoption of such a low technology node (e.g., 65\,nm) in this field remains limited compared to 130/180~nm and above. 

In this proposal, we introduce an 8-channel integrated circuit front-end architecture targetting ToF Computed Tomography (ToF-CT)~\cite{tof}.

\section{Architecture of One Channel}
\label{ARC}
\begin{figure}[tb]
\centerline{\includegraphics[scale=0.2]{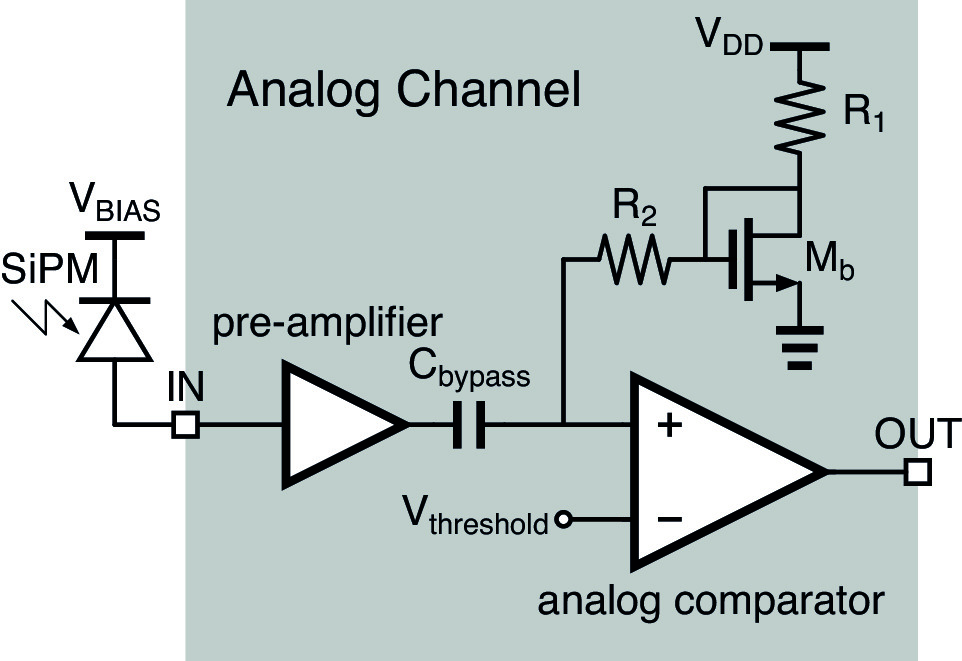}}
\caption{Schematic of one channel including preamplifier and comparator.} 
\label{fig:CH}
\end{figure}

The architecture of an individual channel, depicted in Fig.~\ref{fig:CH}, features a low input impedance amplifier designed to counteract SiPM capacitance and facilitate current-to-voltage conversion, as well as a voltage comparator for gain enhancement. A strategic bypass capacitor between the preamplifier output and the comparator input addresses mismatch concerns from the preamplifier, enabling similar input voltages across all comparators and allowing a uniform threshold to effectively activate each channel. The comparator's input is biased by a diode-connected transistor, $M_b$.  The size of $M_b$ is large to prevent any mismatch effects on the specified node.

\begin{figure}[tb]
\centerline{\includegraphics[scale=0.2]{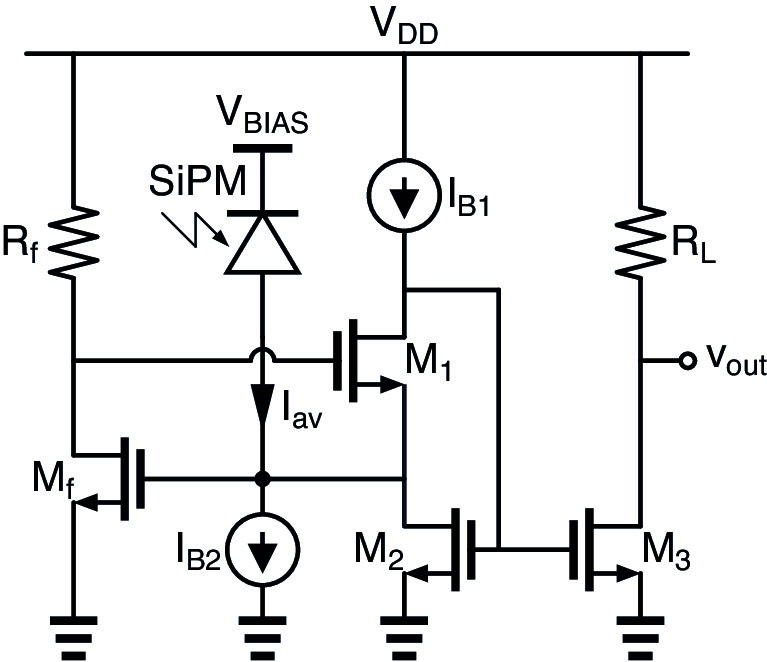}}
\caption{Schematic of the low input impedance preamplifier. $I_{av}$ is the avalanche current of the SiPM.}
\label{fig:amp}
\end{figure}

Fig.~\ref{fig:amp} shows the architecture of the preamplifier. It is primarily based on a flipped voltage follower (FVF)~\cite{b13}, which is biased by a current source $I_{B1}$. Additionally, a new feedback path, composed of $M_f$ and $R_f$, has been added to reduce the input impedance to
\begin{equation}
    R_{in}\approx \frac{1}{g_{m1}g_{m2}R_{B1}(1+g_{mf}R_f)}
\end{equation}
being $R_{B1}$ the equivalent small-signal resistance of $I_{B1}$. 

The size ratio between transistors $M_3$ and $M_2$ significantly influences the input current amplification. 
To enhance this gain, a second bias current source, $I_{B2}<I_{B1}$, has been added. This  approach permits the $M_2$ dc current to remain low, while the size of $M_3$ can be increased without raising power consumption. Simultaneously, $M_1$ maintains a high transconductance (useful for reducing the input impedance) as the current of $I_{B2}$ does not affect its drain current. Obviously, as described in~\cite{b13}, the drain current of $M_2$ (i.e., $I_{B1}-I_{B2}$) cannot be set too low as its transconductance would degrade the input impedance.

Ultimately, the current-to-voltage conversion is facilitated through resistor $R_L$.

\section{Methods and Results}
\label{SIM}

\begin{figure}[htbp]
\centerline{\includegraphics[scale=0.35]{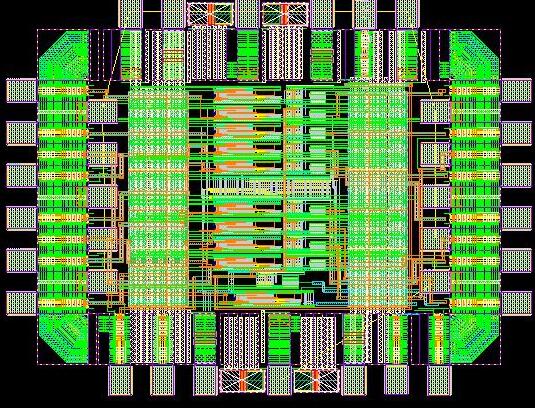}}
\caption{Layout of LUCAS 
including pad rings - each channel has an area of $263\,\textup{µm} \times 40\,\textup{µm}$.}
\label{fig:LAY}
\end{figure}

\begin{figure}[htbp]
\centerline{\includegraphics[scale=0.28]{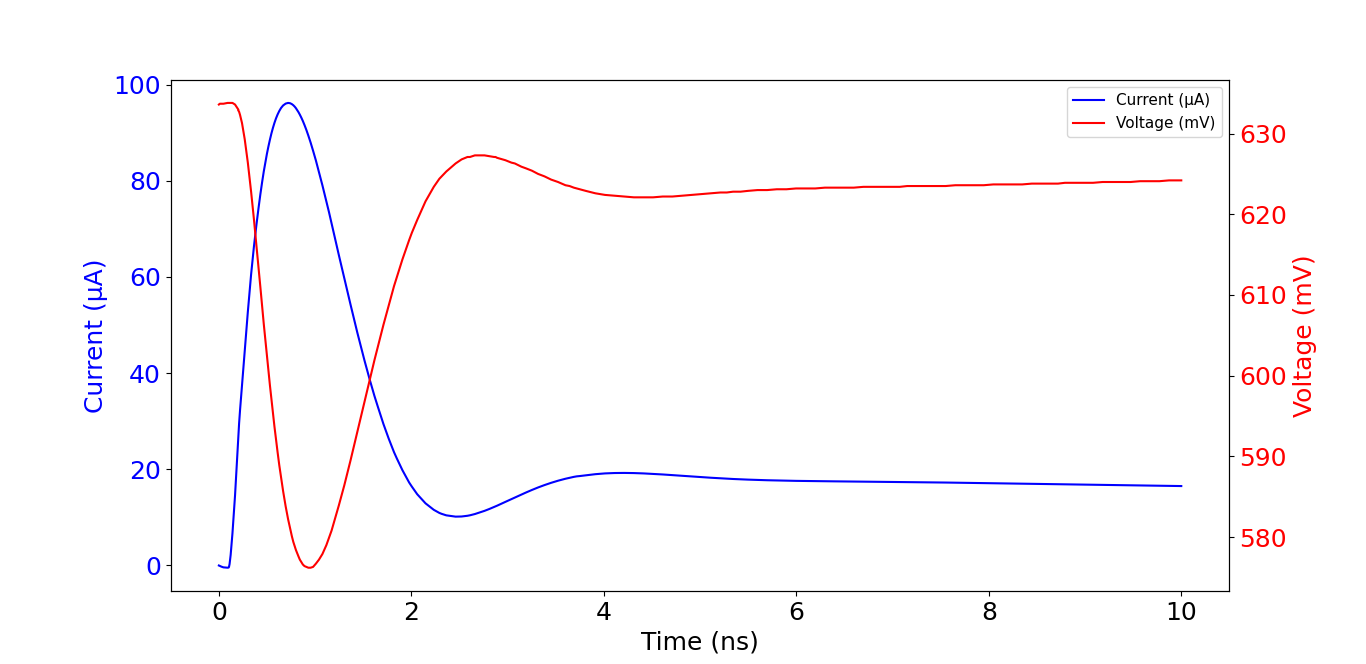}}
\caption{Transient response of the preamplifier for a single photon.}
\label{fig:TRAN}
\end{figure}

The 8-channel front end is simulated in the Cadence Virtuoso environment using a TSMC 65\,nm LP CMOS technology  with a $1.2$ V power supply. The layout of LUCAS is shown in Fig.~\ref{fig:LAY}. Simulation results are carried out at 25$^\circ$C. Each channel's power consumption is $3.2$~mW, the measured jitter is slightly below $40$\,ps FWHM SPTR (calculated based on the equation presented in~\cite{JITTER}), the unity-gain bandwidth is $3.9$\,GHz, and the  DC gain is $560$\,V/I.  The SiPM was modeled using the Verilog-a code from~\cite{b14,b20} that allowed to describe the single-photon arrival. It was adapted to reflect the Broadcom AFBR-S4N44P164,  which is among the targeted devices for the ongoing ToF-CT project at U. Sherbrooke.
Additionally, the impact of wire-bonding is included into the simulations.
The transient response of the preamplifier for single photon detection is shown in Fig.~\ref{fig:TRAN}.
The front end has been submitted for manufacturing and chip samples are expected in August 2023. Experimental measurements corresponding to the above-mentioned simulations will be presented at the conference.

\section{Conclusions}
\label{CON}


This proposal introduces LUCAS, an ultra-low timing jitter analog-front end for SiPM. Its anticipated performances of less than $40$\,ps FWHM SPTR and small physical size make it ideal for integration in a high channel ASIC (64 channels or more) suitable for applications such as a ToF-CT system, which need to support thousands of pixelated channels in a very small volume. This design, coupled with customized SiPM assembly, will reduce routing parasitics and optimize heat extraction, a challenge in very compact systems. The current 8-channel version is suitable for proof-of-concept experimental setups, testing the architecture with different SiPM vendors and improving our usage of simulation models and environments. Future development includes designing an energy channel, integrating a time-to-digital converter (TDC) per channel \cite{CTDemoSystem} and expanding the design to 64/96 channels. Lastly, we anticipate to adapt LUCAS to leverage its ToF performance in the Ultra High Resolution (UHR) PET brain scanner electronics \cite{SherbrookeUHR}, extending LUCAS's impact in the field of nuclear medical imaging.


\begin{thebibliography}{htpb}



\bibitem{b5} L. Buonanno et al. IEEE Transactions on Nuclear Science, 2021.

\bibitem{b9} R. Costanzo et al. IEEE Microwave and Wireless Components Letters, 2018.

\bibitem{TIA} H. Ding et al. IEEE Transactions on Very Large Scale Integration (VLSI), 2023. 

\bibitem{BALUN} S. Pourashraf, et al. IEEE Transactions on Radiation and Plasma Medical Sciences, 2021.


\bibitem{tof} J. Rossignol et al. Physics in Medicine \& Biology, 2020. 

\bibitem{b13} R.G. Carvajal et al. IEEE Transactions on Circuits and Systems I: Regular Papers, 2005. 

\bibitem{JITTER} E.J. Leming,  PhD diss., University of Sussex, 2015.


\bibitem{b14} G. Giustolisi et al. Electronics, 2021. 

\bibitem{b20} G. Giustolisi et al.  European Conference on Circuit Theory and Design (ECCTD), 2013.


\bibitem{CTDemoSystem} D. Roshani et al. IEEE NSS/MIC, 2023, submitted.

\bibitem{SherbrookeUHR}F. Loignon-Houle, et al. IEEE NSS/MIC, 2023, submitted.






\end{thebibliography}
\end{document}